\documentclass[aps,footinbib,showpacs,twocolumn]{revtex4}

\usepackage{amssymb}
\usepackage{graphicx}
\usepackage{dcolumn}
\usepackage{bm}
\usepackage{epsfig}
\begin{document}

\title{On the equation of state of a dense columnar liquid crystal}
\author{H.H. Wensink}
\email{h.h.wensink@chem.uu.nl}
\affiliation{Van 't Hoff Laboratory for Physical and Colloid Chemistry, Debye Institute, Utrecht University, Padualaan 8,
3584 CH Utrecht, The Netherlands}
\pacs{61.30.Cz, 61.30.St, 05.20.Jj}
\date{\today}

\begin{abstract}
An accurate description of a  columnar liquid crystal of hard disks at high packing fractions is presented using an improved  free-volume theory. It is shown that the orientational entropy of the disks in the one-dimensional fluid direction leads to a different high-density scaling pressure compared to the prediction from traditional cell theory. Excellent quantitative agreement is found with recent Monte-Carlo simulation results for various thermodynamic  and structural properties of the columnar state.
\end{abstract}

\maketitle

The lyotropic columnar liquid crystal state, characterized by a two-dimensional hexagonal stacking of columns each with a liquid-like internal structure, has received considerable attention in  recent years  both in experimental colloid science \cite{Brown99,vanderKooijcolumnair} and  computer simulations \cite{Veerman,zhang2}. 
The recently developed colloidal systems of polymer-grafted  gibbsite platelets are known to show a first order phase transition from a spatially homogeneous nematic  to an inhomogeneous columnar phase upon densification \cite{vanderKooijcolumnair}. The columnar signature of the latter is evidenced by its bright Bragg-reflections for visible light and it has  been confirmed on a more rigorous basis using Small-Angle X-ray Scattering (SAXS) measurements \cite{vanderKooijcolumnair}. Colloidal platelets constitute an emerging field of interest
and future efforts can be put into reducing their polydispersity and manipulating columnar textures by means of a magnetic field, both aimed at making high-quality single-domain columnar structures. These may be candidates for the production of e.g. colloidal photonic crystals \cite{colvin}.

\begin{figure*}
\includegraphics[width=14.5cm]{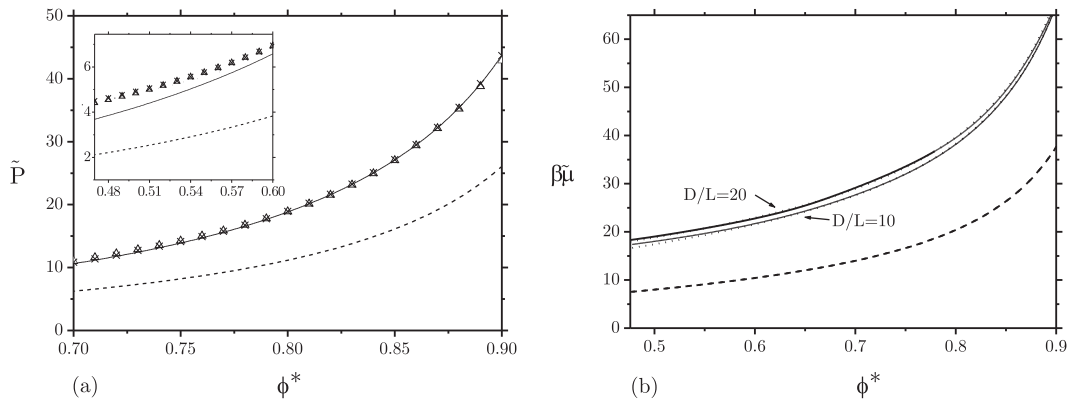}
\caption{(a) Equations of state for the columnar phase. The solid curve is the prediction from the present cell model, the dashed one is the traditional cell pressure, given by Eq. (\ref{tientrad}).
The symbols correspond to the simulation data from  Zhang {\em et al.} \cite{zhang2} for $D/L=10$ (crosses) and $D/L=20$ (triangles). The inset shows the pressure in the dilute regime near the columnar-nematic (``melting") transition at $\phi^{\ast}\approx 0.48$. 
(b) Dimensionless chemical potential $\beta \tilde{\mu}=\beta\mu - \ln\left[ v_{0}/\mathcal{V}\phi_{\text{cp}}\right]$  from simulations (dotted curves) and  cell theory (solid curves). The prediction from the traditional cell model, indicated by the dashed curve, is
independent of $D/L$.}
\end{figure*}


In this Letter we consider a simple but accurate theoretical description inspired by cell theory, which was first applied to spatially ordered liquid crystals by Taylor, Hentschke and Herzfeld \cite{taylor,hentschke}. 
By incorporating the rotational freedom of the disks into the framework we are able to  quantitatively 
account for both thermodynamic and structural properties of the phase along its entire stability range.
The theory is shown to be universal, i.e. independent of the generic shape or the length-to-diameter ratio of the disks.
For this reason, the results are expected to be significant for ongoing experimental studies on colloidal platelet systems. In particular, knowledge of the osmotic pressure
may be useful for analyzing sedimentation experiments \cite{beekschilling} whereas structural predictions may prove helpful for the interpretation of SAXS results on dense systems of platelets \cite{petukhov}.

To describe the properties in the two positionally ordered dimensions  of the columnar liquid crystal we assume the  particles  to be confined within discrete  compartments represented by hexagonal tubular cells which form a close-packed structure according to   the classical free-volume (``cell") theory \cite{eyring,lennardjones}. The disks are allowed to take any position within the cell but they may not share the cell with another particle or penetrate an adjacent one. The cell approach will be combined with an appropriate description of the one-dimensional fluid behaviour along the columnar direction. 
We start from the traditional Tonks model \cite{tonks} applied to a linear fluid of $N$ hard disks with diameter $D$ and thickness $L$ whose centres of mass can move freely on a line with length  $\ell$. Since  the disks are allowed to rotate freely around their centres of mass, the effective excluded thickness $\tilde{L}_{i,j}$ between two adjacent disks $i$ and $j$ is an orientation-dependent quantity, i.e. $\tilde{L}_{i,j}(\Omega_{i},\Omega_{j})>L$, in terms of the solid angle $\Omega$.  We assign $x_{k}$ to the position of particle $k$ on the line and fix the first and last particle at $x_{1}=0$ and $x_{N}=\ell$, respectively. The  configurational integral for this system in the macroscopic limit $L/\ell\rightarrow 0$ is then formally written as
\begin{equation}
Q_{N}=\frac{1}{\mathcal{V}^{N}N!}Q_{N}^{\text{or}}
\left \langle 
\left( \ell - \tilde{L}_{\text{tot}}(\Omega_{1},\ldots,\Omega_{N}) \right)^{N}
\right \rangle _{f(\Omega)}, \label{configdef} 
\end{equation}
with $\mathcal{V}$ the thermal volume  pertaining to the translational and orientational kinetic degrees of freedom. The brackets denote an orientational average according to some unknown  orientation distribution function (ODF) $f(\Omega)$ which is normalized according to 
$\int f(\Omega) d \Omega \equiv 1$. Note that $Q_{N}$ is proportional to an $N$-dimensional free volume with $\tilde{L}_{\text{tot}}$ being the total  occupied length for a given  orientational configuration, expressed in terms of the following sum
\begin{equation}
\tilde{L}_{\text{tot}}(\Omega_{1},\ldots,\Omega_{N})=\sum_{k=1}^{N} \tilde{L}_{k,k+1}(\Omega_{k},\Omega_{k+1}).
\end{equation}
Eq. (\ref{configdef}) is difficult to analyze rigorously so we shall approximate it as follows
\[
Q_{N} \simeq \frac{\ell^{N}}{\mathcal{V}^{N}N!}Q_{N}^{\text{or}}
\left(1- \frac{1}{\ell} \left\langle \tilde{L}_{\text{tot}}(\Omega_{1},\ldots,\Omega_{N}) \right \rangle _{f(\Omega)}\right)^{N}, 
\]
which is assumed to be justified for the strongly aligned orientational configurations we expect in a dense  columnar state.
Further simplification in this respect can be achieved by neglecting the dependency of the excluded length on the azimuthal angle $\varphi $ of the particles. To this end we shall consider an {\em effective} disk thickness, which is determined solely by the polar deflection angle $\theta$ between the particles' symmetry axis and the line unit vector.
For small angles  this quantity is given by
\begin{equation}
\tilde{L}_{\text{eff}}=L \left[ 1+ \frac{1}{2}\frac{D}{L} |\theta | + \mathcal{O}(\theta^{2}) \right], \label{leffe}
\end{equation}
up to leading order in $\theta$.  
The factor `$1/2$' in Eq. (\ref{leffe}) is included explicitly to approximately restore the azimuthal dependency of $\tilde{L}_{\text{eff}}$.
Setting the factor equal to unity yields the {\em maximum} $\tilde{L}_{\text{eff}}$ at fixed polar angles, which is only realized if $\Delta \varphi$ for two adjacent disks is exactly $\pi$. In reality, $\Delta \varphi$ is randomly distributed between 0 and $2\pi$ so that the total excluded length is usually way below its maximum value.
The orientationally averaged total occupied length is approximated by the following mean-field expression:
\begin{equation}
\left \langle \tilde{L}_{\text{tot}} \right \rangle _{f(\Omega)}\simeq  N \left \langle \tilde{L}_{\text{eff}} \right \rangle _{f(\theta)}, 
\end{equation}
where the ODF obeys common uniaxial symmetry and depends only  on the polar angle $\theta$.
The configurational integral then simply becomes
\begin{equation}
Q_{N}=\frac{\ell^{N}}{\mathcal{V}^{N}N!}Q_{N}^{\text{or}}
\left(1-\rho\left[1+\frac{D}{2L}\left\langle |\theta|\right \rangle_{f(\theta)} \right] \right)^{N}, \label{configMF} 
\end{equation}
in terms of the linear density $\rho=NL/\ell$. 
Following Onsager \cite{Onsager} the orientational configurational integral $Q_{N}^{\text{or}}$ reads
\begin{equation}
Q_{N}^{\text{or}}=\exp\left[-N \left\langle \ln [4\pi f(\theta)] \right\rangle_{f(\theta)} \right],
\end{equation}
and the total Helmholtz free energy $\beta F_{\text{fluid}}=-\ln Q_{N}$ of the modified Tonks fluid is given by
a superposition of the ideal, orientational and configurational entropic contributions:
\begin{eqnarray}
\frac{\beta F_{\text{fluid}}}{N} &=& \ln [\rho \mathcal{V}]+\int f(\theta)\ln[4\pi f(\theta)] d\Omega \nonumber \\
&&-\ln\left[ 1-\rho\left(1+ \frac{D}{2L} \int  f(\theta)|\theta| d\Omega\right)   \right], \label{tiensuper}
\end{eqnarray}
with $\beta=1/k_{B}T$.
The thermodynamic equilibrium ODF can be obtained by a functional  minimization of the free energy with respect to the orientational distribution under the normalization restriction.  After some algebra we arrive at the following {\em closed} expression for the normalized ODF: 
\begin{equation}
f(\theta)=\frac{\alpha ^{2}}{4\pi} \exp[-\alpha |\theta|], \label{tienODF}
\end{equation}
where $\alpha$ depends on the density and the disk aspect ratio $D/L$ via
 \begin{equation}
\alpha=\frac{3}{2}\frac{D}{L}\left(\frac{\rho}{1-\rho} \right), \label{tienalfa}
\end{equation} 
Since $\alpha\gg 1$ for sufficiently anisometric disks ($D/L\gg 1$) the ODF is sharply peaked around $\theta=0$, as we expect.  The divergence of $\alpha$ at close packing ($\rho = 1$) indicates that the disks are forced to orient parallel to the director in this limit.
Inserting Eq. (\ref{tienODF}) and straightforward integration yields an explicit free energy  in terms of the density $\rho$. Taking the standard derivative with respect to the density then gives the (dimensionless) pressure $\beta P L$ of the one-dimensional fluid:  
\begin{equation}
\beta P L  = 3\rho / (1-\rho), \label{modtonks}
\end{equation} 
which is exactly three times the original Tonks pressure \cite{tonks}. This result implies that the
orientational confinement of the disks inside the columns gives rise to an additional entropic contribution  $2\rho/(1-\rho)$ to the total pressure. Note that the derivation of Eq. (\ref{modtonks})  closely resembles the origin of the quasi-ideal pressure of the nematic state $ \beta P \sim 3 N/V$ within the Gaussian scaling Ansatz for the ODF \cite{OdijkLC,OdijkLekkerkerker}, although both have a  different physical basis.

The description of the columnar phase in the two positionally ordered dimensions is analogous to that of a 2-D (hexagonally) ordered configuration of $N$ disks. According to the non-correlated  version of the cell model the configurational integral of the $N$-particle system can be approximated by \cite{kirkwoodfreevol}:
\[
Q_{N}^{\text{cell}} =\int d{\bf r}^{N} \exp[\beta U({\bf r}^{N})]
\simeq   \left(\int d{\bf r}\exp \left[-\frac{\beta}{2} u_{\text{cell}}^{\text{nn}}({\bf r}) \right] \right)^{N}
\]
where $u_{\text{cell}}^{\text{nn}}({\bf r})$ is the potential energy between the particle and its nearest neighbours. For hard interactions the second phase space integral is simply the (2-D) {\em free volume} of the particle in the cell.  Assuming that the nearest neighbours constitute a perfect hexagonal cage, this free volume is  given by $V_{\text{free}}=\sqrt{3}(\Delta_{c}-D)^{2}/2$ with $\Delta_{c}$ the nearest neighbour distance. The configurational integral then becomes (ignoring all irrelevant contributions)
\begin{equation}
Q_{N}^{\text{cell}}\simeq  \left(V_{\text{free}}\right)^{N} 
\propto  \left( 1-\bar{\Delta}_{c}^{-1}\right)^{2N},
\end{equation}
in terms of the dimensionless spacing $\bar{\Delta}_{c}=\Delta_{c}/D$. 
Applying the condition of single-occupancy (i.e. one array of disks per column) we can use $\bar{\Delta}_{c}$ to relate the linear density $\rho$ to the three-dimensional volume fraction via
\begin{equation}
\rho=\phi^{\ast} \bar{\Delta}_{c}^{2}, \label{tienrhophi}   
\end{equation}
with $\phi^{\ast}=\phi/\phi_{\text{cp}}$ the volume fraction normalized to   its close-packing value $\phi_{\text{cp}}=\pi/2\sqrt{3}$.

The total excess Helmholtz free energy of the columnar state is obtained by adding the cell contribution to the Tonks excess free energy. Omitting all  constant terms we arrive at
\begin{equation}
\frac{\beta F^{\text{ex}}_{\text{tot}}}{N} \sim  -2 \ln \left[ \frac{\rho}{1-\rho}\right ]-\ln\left[1-\rho \right]-2 \ln \left [ 1-\bar{\Delta}_{c}^{-1} \right], \label{tienexcessfree}	 
\end{equation}
representing  the orientational, `Tonks' and cell contributions, respectively. Inserting Eq. (\ref{tienrhophi}) and minimizing the free energy \footnote{By definition, the ideal free energy ($\sim\ln \rho$) does not depend on the cell spacing and therefore drops out of the free energy minimization.}  with respect to the cell spacing $\bar{\Delta}_{c}$ yields a third-order consistency equation with complicated solutions. Expanding the physical solution near close-packing in terms of $1-\phi^{\ast}$ up to leading order yields
\begin{equation}
\bar{\Delta}_{c} =1+\frac{1}{5}(1-\phi^{\ast}) + \mathcal{O}[(1-\phi^{\ast})^{2}]. \label{inter}
\end{equation}
Substituting its closed form  into the free energy Eq. (\ref{tienexcessfree}) and taking the appropriate derivative gives a similar expansion for the total dimensionless pressure $\tilde{P}=\beta P v_{0}/\phi_{\text{cp}}$ (with $v_{0}$ the disk volume), i.e.
\begin{equation}
\tilde{P} = \frac{5}{1-\phi^{\ast}} + 6.4 + 1.128 (1-\phi^{\ast})+\mathcal{O}[(1-\phi^{\ast})^{2}], \label{tiendruk}
\end{equation}
indicating that the high-density scaling pressure, i.e. the leading order contribution, is essentially different from the classical cell prediction $3/(1-\phi^{\ast})$  for hard spheres \cite{wood,aldersimul}. The latter result is completely analogous to  our result for {\em perfectly aligned} disks and can be reproduced directly from Eq. (\ref{tienexcessfree}) by omitting the orientational contribution given by the first term. The ``traditional" cell pressure is given by
\begin{equation}
\tilde{P}\equiv \frac{\phi^{\ast}}{1-(\phi^{\ast})^{1/3}}= \frac{3}{1-\phi^{\ast}} + 4 +\mathcal{O}(1-\phi^{\ast}). \label{tientrad}
\end{equation}
In Fig. 1(a) we have plotted the abovementioned cell equations of state, i.e. Eq. (\ref{tientrad}) and the closed-form analog of Eq. (\ref{tiendruk}), along with the ones obtained from Monte-Carlo simulations on cut-spheres by Zhang {\em et al.} \cite{zhang2}. 
The quantitative agreement between the present cell description and the simulations  significantly improves upon densification. Above $\phi^{\ast}\approx 0.8$ our prediction is found to match the simulations within 1\%. Obviously, the discrepancy is much larger in the dilute regime ($\phi^{\ast}<0.6$) where the cell model, at least the simplest version considered here, is no longer quantitatively reliable.

\begin{figure}
\includegraphics[width=7cm]{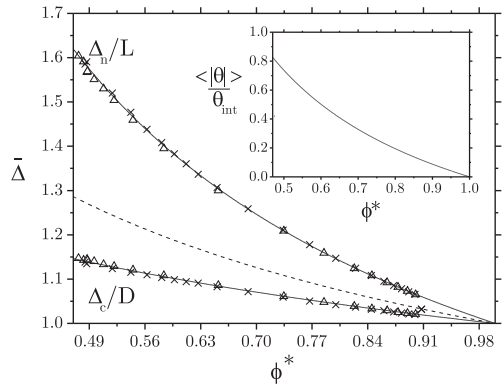}
\caption{ Normalized inter- and intracolumnar spacings, $\Delta_{c}/D$ and $\Delta_{n}/L$ respectively, as a function of $\phi^{\ast}$. Solid lines are theoretical predictions, the symbols follow from simulations for $D/L=10$ (crosses) and $D/L=20$ (triangles). The dotted curve follows from the traditional cell model and denotes {\em both} spacings. Inset: Average polar angle relative to the internal angle ($\theta_{\text{int}}\sim L/D$) plotted versus $\phi^{\ast}$. }
\end{figure}

The chemical potentials are depicted in Fig. 1(b). These are readily obtained from the absolute free energy and pressure of the cell model via the Legendre transform $\mu=(F+PV)/N$  and by means of a Gibbs-Duhem integration of the pressure fits obtained from the simulation data \cite{zhang2}.
Again, close agreement is found between both. The values at melting 
are found to differ only by $\Delta \tilde{\mu} \approx 0.3 k_{B}T$ and $0.6k_{B}T$ for $D/L=20$ and $D/L=10$ respectively, indicating that the cell prediction is  surprisingly efficacious  throughout the $\phi^{\ast}$-range. The ``splitting" of the curves for both aspect ratios is simply due to the orientational entropy [second term in Eq. (\ref{tiensuper})], $\beta F_{\text{or}}/N=2\ln\alpha-2$,
which depends explicitly on $D/L$ via Eq. (\ref{tienalfa}) and therefore gives rise to a different intercept at the melting volume  fraction.

Let us now focus on the spacings  between the columns $\bar{\Delta}_{c}$ and the average disk spacing inside the columns $\bar{\Delta}_{n}\equiv \Delta_{n}/L=\rho^{-1}$, which follows directly from
Eqs. (\ref{tienrhophi}) and (\ref{inter}):  
\begin{equation}
\bar{\Delta}_{n}= \left[\phi^{\ast}\left(1 +\frac{2}{5}(1-\phi^{\ast}) \right)\right ]^{-1} + \mathcal{O}[(1-\phi^{\ast})^{2}], \label{intra}
\end{equation}
Comparing with Eq. (\ref{inter}) we see that the intracolumnar distance $\bar{\Delta}_{n}$ between the disks grows faster than the
intercolumnar one $\bar{\Delta}_{c}$.  The expansion of the columnar structure is therefore  {\em anisometric}. This behaviour is quite different from the classical model for which $\bar{\Delta}_{c}\equiv \bar{\Delta}_{n}=\phi^{\ast -1/3}$ indicating an {\em isometric} expansion upon lowering $\phi^{\ast}$.
In Fig. 2 the predicted spacings, given by the closed-form analogs of Eqs. (\ref{inter}) and (\ref{intra}), are compared with simulation results. Unlike the pressure, the prediction for the columnar spacings remains surprisingly accurate even in the regime close to the columnar-nematic transition. Quantitatively, theory and simulation are found to agree  within 1\% over the entire columnar stability range.

To assess the degree of orientational order in the  dense columnar state we consider the ratio of the
average polar angle $\left \langle |\theta| \right \rangle $ to the {\em internal} angle $\theta_{\text{int}}\sim L/D$,  i.e.  $\left \langle |\theta| \right \rangle / \theta_{\text{int}}=2D/L\alpha$. Substituting Eqs. (\ref{tienalfa}) and (\ref{intra}) reveals a simple proportionality with respect to the intracolumnar spacing, i.e. 
$\left \langle |\theta| \right \rangle /  \theta_{\text{int}}\simeq  (4/3)(\bar{\Delta}_{n}-1)$.
This tells us that the disks are only marginally perturbed away from their parallel orientations since the average ``off-parallel" deflection angle does not exceed the internal angle of the disk.
The full result, included in Fig. 2,  explicitly shows that this situation remains  up to  the columnar-nematic transition, located around $\phi^{\ast}\approx 0.48$.  From this we conclude  that the orientational freedom of the disks is extremely small throughout the {\em entire} columnar stability regime. Moreover, the dominance of near-parallel configurations gives an {\em a posteriori} justification for the mean-field asymptotic analysis presented here. 

In conclusion, we have constructed a modified  cell theory for the columnar state by explicitly accounting for the rotational freedom of the disks. Our  approach  constitutes a significant quantitative improvement over the traditional one which, contrary to the case of a hard-sphere FCC-crystal \cite{aldersimul},  appears to be inappropriate for a columnar liquid crystal even in the regime near close-packing.  The theory does not only quantitatively predict the thermodynamics of the columnar state, as evidenced by the pressure and chemical potential, but also  structural features in terms of the characteristic columnar spacings.

Future work could be aimed at applying the present approach to a dense smectic-A phase of hard spherocylinders by considering a two-dimensional fluid of rotating rods \cite{taylor}. Although the theory is expected to be less successful here, it would be intriguing to verify its quantitative merits for a lyotropic smectic phase. 
Furthermore, the cell description could be refined in the regime close to the melting transition by introducing more advanced cell theories which include e.g.  cooperative motion of the columns \cite{aldercoop}.
The fact that the simulation pressures in Fig. 1(a) remain insensitive to the aspect ratio throughout the entire columnar stability range is surprising and  supports the notion that the columnar phase is dominated by cell-behaviour up to the melting transition, albeit in a more sophisticated fashion than we described here.  

I am grateful to Henk Lekkerkerker, Andrei Petukhov and Gert Jan Vroege for stimulating discussions and a critical reading of the manuscript.  Jeroen van Duijneveldt and Shu-Dong Zhang are kindly thanked for providing their simulation data.

\end{document}